\documentclass[pre,twocolumn,aps,amssymb,footinbib,showpacs]{revtex4}
\usepackage{amssymb}
\usepackage{amsmath}
\usepackage{times}
\usepackage{latexsym}
\usepackage{graphicx}
\usepackage{color}
\usepackage{bm}
\usepackage{subfigure}
\begin{document}
\title{Soliton dynamics of an atomic spinor condensate on a Ring Lattice }
\author{Indubala I. Satija$^{1}$, Carlos L. Pando L.$^{2}$ and Eite Tiesinga$^{3}$}
 \affiliation{$^{1}$School of Physics, Astronomy and Computational Sciences, George Mason University,
 Fairfax, VA 22030, USA}
\affiliation{$^{2}$IFUAP, Universidad Autonoma de Puebla, Apdo, Postal J-48, Puebla 72570, Mexico}
\affiliation{$^{3}$Joint Quantum Institute, National Institute of
Standards and Technology and the University of Maryland, 100 Bureau Drive,
Stop 8423, Gaithersburg, MD 20899-8423, USA}
\date{\today}

\begin{abstract}
We study the dynamics of macroscopically-coherent matter waves of an
ultra-cold atomic spin-$1$ or spinor condensate on a ring lattice of
six sites and demonstrate  a novel type of spatio-temporal internal
Josephson effect.  Using a discrete solitary mode of uncoupled spin
components as an initial condition, the time evolution of this many-body
system is found to be characterized by two dominant frequencies leading
to quasiperiodic dynamics at various sites.  The dynamics of {\it
spatially-averaged} and {\it spin-averaged} degrees of freedom, however,
is periodic enabling an unique identification of the two frequencies.
By increasing the spin-dependent atom-atom interaction strength we
observe a resonance state, where the ratio of the two frequencies is
a characteristic integer multiple  and the spin-and-spatial degrees
of freedom oscillate in ``unison''.  Crucially, this resonant state is
found to signal the onset to chaotic dynamics characterized by a broad
band spectrum.  In a ferromagnetic spinor condensate with attractive
spin-dependent interactions, the resonance is accompanied by a transition
from oscillatory- to rotational-type dynamics as the time evolution of
the relative phase of the matter wave of the individual spin projections
changes from bounded to unbounded.
\end{abstract}


\maketitle

\section{Introduction}

Spinor condensates are atomic Bose-Einstein condensates (BEC)
with an internal spin degree of freedom that combine magnetism with
condensation.  Examples are optically-trapped spinor condensates of atomic
rubidium-87\cite{Kuwamoto2004,Chang2004,Schmaljohann2004,Guzman2011},
sodium\cite{Black2007,Liu2009}, or chromium \cite{Pasquiou2011}
with either spin or angular momentum $f=1$, 2, or 3 condensates with a
three-, five-, or seven-component vector order parameter.  In contrast
to mixtures of two or more atomic states\cite{Myatt1997} or mixtures of
several atomic species, spin-changing collisions in spinor condensates
permit coherent dynamics among the hyperfine states.  In a typical
process for two $f=1$ atoms, one in spin component $m=-1$ and one
in $m=+1$, can reversibly scatter into two atoms with spin component
$m=0$, which conserves the global magnetization of the condensate. This
coherent spin mixing leads, nevertheless, to oscillations of the spin
populations, and is an analogue of Josephson oscillations in ultra-cold
atoms\cite{Pu1999,Zhang2005,Santos2006,Barnett2010,Yukawa2012}.
Hence, just as collisional interactions allow for a single-component
condensate to be spatially coherent, spin-changing collisions, driven
by internal interactions, allow coherence among internal degrees of
freedom.  A positive or negative sign of the strength of the spin-changing
interaction determines whether the systems behave anti-ferromagnetic or
ferromagnetic, respectively \cite{Zhang2005,Barnett2006}.
Spinor physics has also been studied in optical lattices with exactly two atoms
per lattice site \cite{Widera2005}.

In view of the nonlinear nature due to interparticle interactions and high
degree of control in the experiments, BEC systems are ideal systems for
visualizing a wide variety of nonlinear phenomena. This includes solitary
waves, the localized nonlinear traveling waves that retain their shape,
size and speed during propagation \cite{soliton}.
Experimental realization of solitons in a homogeneous single component
BEC systems is a hallmark of the quantum coherence associated with many
body systems.\cite{Burger1999,Denschlag2000,Mishmash2009} There has also
been theoretical studies of solitons in a homogeneous spin-1 condensate
\cite{Wadati2006,Ieda2006}. Furthermore, theoretical and experimental
studies of double-well bosonic Josephson junctions have unveiled novel
phenomena such as broken symmetry macroscopic quantum self-trapping
\cite{Satija2009}and $\pi$-modes\cite{Smerzi1997,Albiez2005,Mahmud2005}
as well as symmetry restored swapping modes\cite{Satija2009}.
In addition, numerical investigation of BEC systems, spatially
separated into a ring lattice have demonstrated chaotic dynamics,
deterministic dynamics with sensitive dependence on initial
conditions.\cite{Pando2004,Buonsante2007,Pando2009}

\begin{figure}
\includegraphics[scale=0.24,clip]{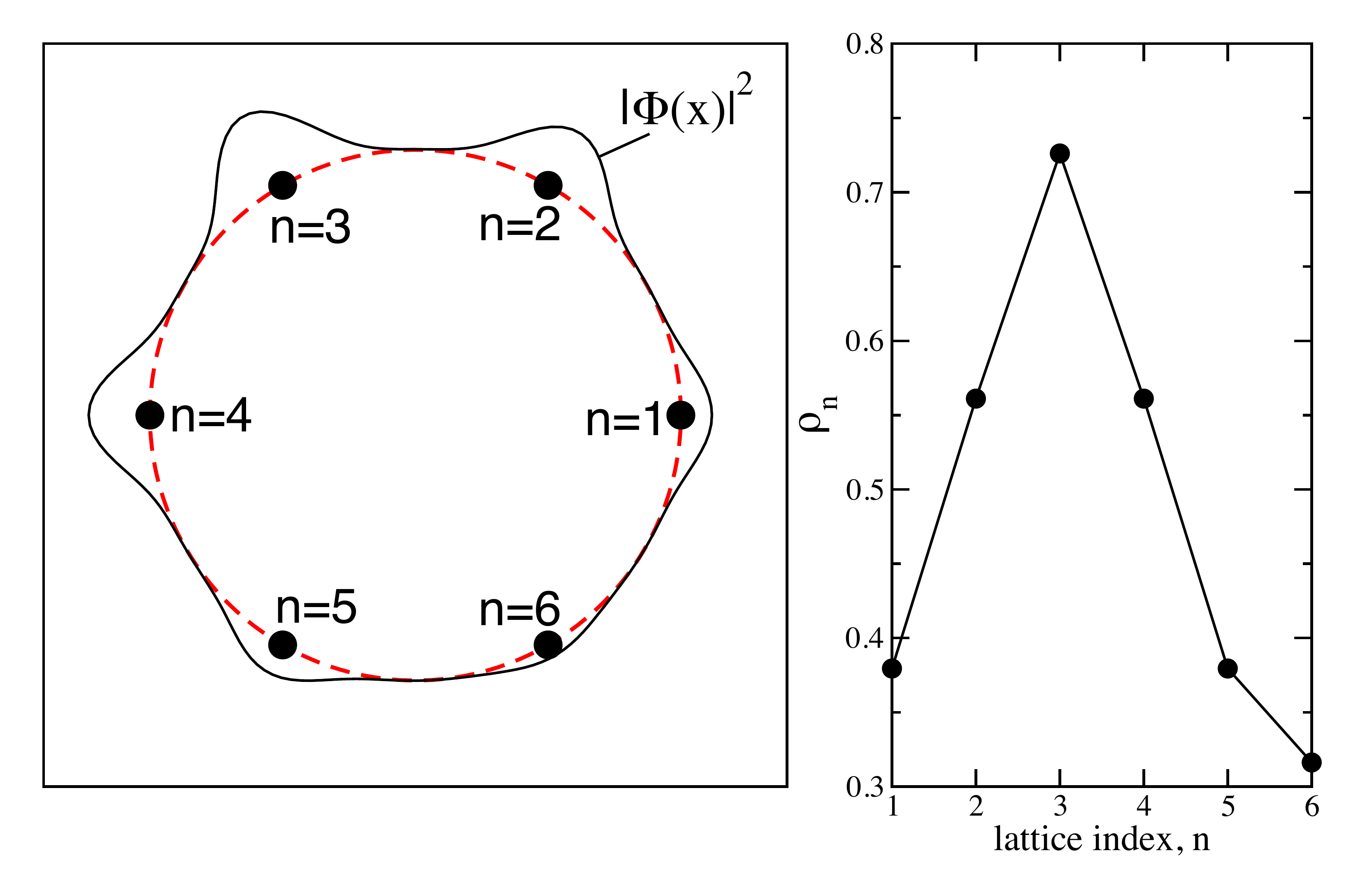}
\caption{(color online) 
Panel a) shows a schematic of  a ring lattice of six sites (red dashed
circle with labeled markers) containing a $f=1$ spinor condensate.
The solid black line represents the superfluid population along the
ring as a polar graph.  Panel b) shows the initial fractional population,
$\rho_n$, as a function of site index $n$ used in our dynamical
simulations.  The distribution corresponds to an excited (period-6)
solution of the DNLSE.  For each site the population of the three magnetic
sublevels of the spinor is the same.
  }
\label{F0}
\end{figure}

In this paper, we explore the quantum coherent time evolution of a spin-1
BEC that is spatially separated into six weakly-coupled sites arranged to form
a ring geometry, see Fig.~\ref{F0}.  
For a large number of atoms
such a system is well represented by a three-component wavefunction
or order parameter, $\vec \Phi(\vec x,t)$, that satisfies a nonlinear Gross-Pitaevskii
equation\cite{Pu1999,Zhang2005}.  Here, we restrict our calculations to
the case with no external magnetic field. More importantly, we assume that
all sites and spin components have at all times the same identical
localized spatial-mode function, $\phi(\vec x)$.  That is, for the $m^{\rm th}$-component
of $\vec \Phi(\vec x,t)$ we have 
\begin{equation}
    \Phi_m(\vec x,t) =  \sum_{n=1}^L \psi_n^m(t) \phi(\vec x-\vec x_n) \,,
\label{eq_SMA}
\end{equation}
where $L$ is the number of sites, $\vec x_n$ is the center
of site $n$, and the dimensionless $\psi_n^m(t)$ are complex time-dependent
amplitudes.  
The overlap between spatial-mode
functions at different sites leads to tunneling between neighboring
sites.

More advanced numerical modeling uses time varying mode functions
\cite{Anker2005,Alon2007} or directly solves for the three-dimensional
Gross-Pitaevskii equation\cite{Denschlag2000}.  As a first study in the soliton
dynamics of a spinor condensate we believe that our simplified approach
is justified. It shows the generalization of the spinor Josephson effect
as well as the route to chaos in its clearest form.
It should be noted that the ring lattice with less than three sites is an
integrable system.  BEC on a lattice with three or more than three sites
exhibits amazing degree of complexity including coexistence of regular
and irregular behavior that is known to accompany chaotic dynamics.
In our detailed numerical exploration of ring lattices with different
numbers of sites, the key characteristics of the time evolution as
discussed in this paper are independent of the number of sites. Here,
we will present our results for a ring lattice of six sites only.

Our initial state is a discrete solitary wave that is an excited
eigenstate of the discrete nonlinear Schr\"odinger equation (DNLSE)
satisfied by the $\psi_n^m(t)$ in the absence of spin-changing interactions.
The spin-changing interaction then induces oscillatory dynamics, an internal
Josephson effect where different spin components evolve with different
amplitudes and phases.  In reality the spin-changing Hamiltonian can not
be turned off and this state needs to be engineered by a combination
of resonant electro-magnetic radiation that controllably creates superpositions of
$m$-states \cite{Liu2009} and off-resonant spatially-dependent light
forces that induce solitons \cite{Burger1999,Denschlag2000}.

In this article we show that for weak spin-changing interactions the
dynamics on the ring lattice is dominated by two frequencies.
Interestingly, one of these frequencies, which we denote as $\omega$,
describes the periodic behavior of {\it spatially-averaged} spin
populations. The second frequency,
denoted by $\Omega$, describes the dynamics of the site-dependent
population summed over the three spin components. In other words,
the dynamics of collective or averaged coordinates is pendulum-like.
As we discuss below, this is somewhat surprising as the spatial profiles
of the three spin components evolve differently and the dynamics on the
ring should deviate from that for a ``simple'' trapped spinor condensate.
The single-mode approximation for the latter system has been shown
to lead to pendulum-type physics \cite{Barnett2010,Zhang2005}.
Our study suggests that the origin of this ``regular'' global dynamics
has its roots in the strong correlations among different lattice
sites and spin projections and is not rooted in thermal
averaging over many degrees of freedom.
We note that  $\omega$ is solely determined by the spin-changing
interaction, whereas $\Omega$ shows a rather weak dependence. We find
that $\Omega$ is mainly controlled by the size of the ring.

By increasing the absolute value of the spin-dependent interaction
strength, the two frequencies can be mode locked.  This resonance
condition is found to describe the onset to chaotic dynamics. In
other words, the spinor condensate exhibits a {\it  transition from
quasi-periodic dynamics to chaotic dynamics where at the onset
to transition, all local and global degrees of freedom oscillate
in unison}.  Consequently, the initial soliton profile reappears
periodically, providing a unique demonstration of a novel-type of
spatio-temporal internal Josephson effect.  Furthermore, in contrast
to the anti-ferromagnetic case, in the ferromagnetic condensate, the
resonance is accompanied by a transition from bounded to unbounded
dynamics.  In the anti-ferromagnetic case, the dynamics remains bounded
and oscillatory. Nevertheless, beyond the resonance additional frequencies
appear.

For sake of simplicity, we restrict our simulations to zero magnetization.
This is enforced by choosing initial conditions where the $m=+1$ and
$m=-1$ components have same initial wave function. In fact, we choose an
initial state where all three components have the same initial wave function.
An alternate initial state with zero magnetization and contains mostly
$m=\pm1$ atoms lead to similar results. It will not be described here.

In section \ref{sec_mf} we review the mean-field equations that describe
the evolution of the spinor condensate on a lattice as well as the
underlying single-mode approximation approximation on each lattice
site. The latter has been shown to provide a reasonable description for
the continuum system. For our ring lattice, the SMA suggests collective
coordinates to describe the global dynamics. The initial state is
described in Sec.~\ref{sec_initial}.  In section \ref{sec_weakspin},
we show numerical simulations for weak positive, anti-ferromagnetic
spin-dependent integrations and show that the dynamics, although complex,
is dominated by two frequencies.  In section \ref{sec_resonance}, we
discuss the resonance condition that occurs when the spin-dependent
interaction strength is increased. At resonance we observe an onset
to chaos.  Section \ref{sec_ferro} briefly illustrates the ferromagnetic
case.  We conclude in Sec.~\ref{sec_discussion}.

\section{Mean-field Equations for a Spinor Condensate and 
Collective Coordinates} \label{sec_mf}

An atomic spinor condensate with large atom number is described within
mean-field theory by a complex vector order parameter or condensate
wave function $\Psi({\vec x},t)$, whose evolution is governed by a
multi-component Gross-Pitaevskii equation \cite{Pu1999,Zhang2005}.
We will assume that the order parameter can be approximated by
Eq.~\ref{eq_SMA}, where a single time-independent mode function,
$\phi({\vec x})$, determines the spatial dependence in each well.
We denote this by the $L$-site single-mode approximation (L-SMA) in
analogy to the single-mode approximation (SMA) for a spinor condensate
in a dipole trap.

The interactions between two spin-1 atoms have a
spin-independent and a spin-dependent or (spin-changing) contribution. Within a mean-field
theory and the L-SMA their strength is given by $c_q=4\pi \hbar^2
g_q/(2\mu)  \int d{\vec x} |\phi({\vec x})|^4/\int d{\vec x} |\phi({\vec x})|^2$ with $q=0$ and 2 for the
spin-independent and spin-dependent contribution, respectively. Here
$\mu$ is the reduced mass for two atoms and $\hbar$ is the reduced
Planck constant. The lengths $g_q$ are $g_0=(a_0+2a_2)/3$ and $g_2 =
(a_2-a_0)/3$, where $a_0$ and $a_2$ are scattering lengths for $s$-wave
collisions of two $f=1$ bosons with total angular momentum $F=0$
and $2$, respectively.  For stable condensed gases we require that
$c_0>0$.  We note that $c_2 > 0$ for anti-ferromagnetic Na and $c_2 <
0$ for ferromagnetic $^{87}$Rb.

The dynamics of the $\psi^m_n(t)$ are governed by the DNLSE 
\begin{eqnarray}
\lefteqn{i \hbar\dot{\psi}^{-1}_n  =  -J (\psi^{-1}_{n-1}+\psi^{-1}_{n+1})}
     \label{eq_one} \\
 &&\quad\quad\quad +(c_0+c_2) (|\psi^{-1}_n|^2 +|\psi^0_n|^2)\psi^{-1}_n \nonumber \\
&&\quad\quad\quad\quad +(c_0-c_2) |\psi^{1}_n|^2 \psi_n^{-1}+c_2(\psi^0_n)^2 (\psi^{1}_n)^*
   \nonumber\\
\lefteqn{i \hbar\dot{\psi}^0_n  =  -J (\psi^0_{n-1}+\psi^0_{n+1})+c_0|\psi^0_n|^2 \psi^0_n} 
 \\
&&\quad +(c_0+c_2) (|\psi^1_n|^2 +|\psi^{-1}_n|^2)\psi^0_n +2c_2 \psi^1_n \psi^{-1}_n (\psi^0_n)^*  \nonumber\\
\lefteqn{i\hbar \dot{\psi}^1_n  = -J (\psi^1_{n-1}+\psi^1_{n+1})+(c_0+c_2) (|\psi^1_n|^2+|\psi^0_n|^2)\psi^1_n} \nonumber\\
&&\quad\quad\quad +(c_0-c_2) |\psi^{-1}_n|^2 \psi_n^1
+c_2(\psi^0_n)^2 (\psi^{-1}_n)^*
\,,  \label{eq_three}
\end{eqnarray}
where $J$ is the positive site-to-site tunneling energy and we use
periodic boundary conditions.  In the absence of the last term on the
right hand side of Eqs.~\ref{eq_one}-\ref{eq_three}, the set of equations
describes a three-species condensate. Spin-changing terms make a spinor
condensate unique as they induce population oscillations between $m$
levels.  Equations~\ref{eq_one}-\ref{eq_three} conserve total atom
number and magnetization.  In other words, $\sum_{nm} |\psi^m_n(t)|^2$
and $\sum_{nm} m|\psi^m_n(t)|^2$ are conserved.

We will monitor the local population of each spin state as well as
global or collective coordinates, such as spatially- and spin-averaged
population and phases.  It is therefore convenient to define $\psi_n^m(t)=
\sqrt{\rho_n^m(t)}\exp[i\phi_n^m(t)]$ with populations $\rho_n^m(t)$
and phases $\phi_n^m(t)$.  Following Refs.~\cite{Pu1999,Zhang2005}
for a single-mode spinor condensate natural local canonical coordinates are
\begin{equation}
Z_n = \rho^0_n-(\rho^{+1}_n+\rho^{-1}_n) \quad{\rm and} \quad 
\gamma_n = \phi^{-1}_n+\phi^{1}_n-2\phi^0_n  \,.
 \label{eq_can}
\end{equation}
as well as globally-averaged coordinates $Z=\sum_n Z_n/L$ and $\gamma=\sum_n\gamma_n/L$.
Throughout this article we call $\gamma_n$ and $\gamma$ spinor phases.
Other useful population averages are
\begin{equation}
\rho_m=\frac{1}{L}\sum_{n=1}^L \rho_n^m \quad {\rm and}\quad
\sigma_n =\frac{1}{3}\sum_{m=-1}^1 \rho_n^m \,.
\label{collect}
\end{equation}

In a dipole trap, or equivalently for $L=1$, a simple spinor model is given by the SMA.
The canonical variables $Z$ and $\gamma$ then
satisfy the pair of equations
\begin{equation}
\dot{Z} = \frac{c_2}{\hbar} (1-Z^2) \sin\gamma \quad{\rm and}\quad
\dot{\gamma}= -2\frac{c_2}{\hbar} Z (1+\cos\gamma) \,,
\label{Heqn}
\end{equation}
which are independent of the spin-independent interaction with strength $c_0$.
For small $Z$ and $\gamma$ the dynamics are harmonic. In general, however,
these coupled equations describe a nonlinear pendulum whose length depends
upon the momentum. 

For a $L>2$ ring of lattice sites
the collective variables $Z$ and $\gamma$ satisfy 
\begin{eqnarray}
\dot{Z} &=& \frac{c_2}{\hbar} \sum_n (\sigma^2_n -Z_n^2) \sin\gamma_n\\
\dot{\gamma}&=&- 2\frac{c_2}{\hbar} \sum_n Z_n (1+\cos\gamma_n )+\frac{J}{\hbar} W
\label{rhoeqn}
\end{eqnarray}
where $W={\rm Re}[\sum_n (2\psi^0_{n+1}/\psi^0_n-\psi^{+1}_{n+1}/\psi^{+1}_n-\psi^{-1}_{n+1}/\psi^{-1}_n)]$.
The spin-independent interaction strength $c_0$ does not explicitly appear
in these equations.  As already mentioned in the introduction and further
discussed in next section, this suggests that the spatially-averaged $Z$
and $\gamma$ will exhibit periodic oscillations, with frequency $\omega$,
that are solely determined by the spin-changing interaction.
Finally, the spin-averaged populations at each site satisfy
\begin{equation}
\dot{\sigma}_n=-\frac{J}{\hbar}\, {\rm Im}\left(\sum^1_{m=-1} (\psi_n^m)^* [\psi^m_{n+1}+\psi^m_{n-1}]\right)\,,
\end{equation}
which does not explicitly depend on $c_2$.  Hence, we expect that the
$\sigma_n$ oscillate periodically, characterized by frequency
$\Omega$ and its harmonics.

\begin{figure}
\includegraphics[scale=0.43,trim=0 150 0 50,clip]{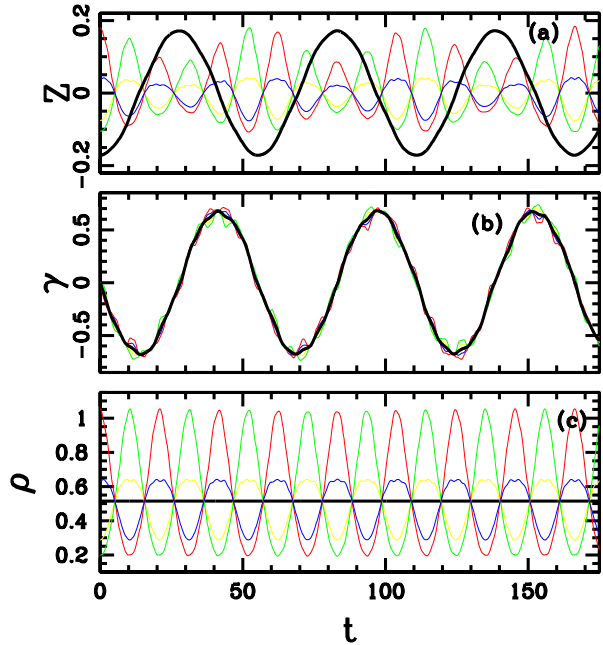}
\caption{(color online) Time dynamics of an anti-ferromagnetic spinor soliton on a six-site
ring for several spin- and/or spatially-averaged degrees of freedom
assuming a small positive spin-changing interaction energy.  Time is
in units of $\hbar/J$, where $J$ is the tunneling energy between the
sites. Calculations are performed for $c_2=0.1J$ and $c_0=J$.  Panel a)
shows the dynamics of the spatially-averaged spinor population $Z$
(black line) and the spinor population $Z_n$ for the individual sites
$n$ (colored lines). The symmetry of the initial soliton around $n=3$
implies that only four of the six sites have a distinct time evolution.
Panel b) shows the spinor phases ${\gamma}$ (black line) and $\gamma_n$
(colored lines).  Finally, panel c) shows the site-specific population
$\sigma_n$, averaged over the three spin components.  }

\label{spin6}
\end{figure}

\begin{figure}
\includegraphics[scale=0.43,clip]{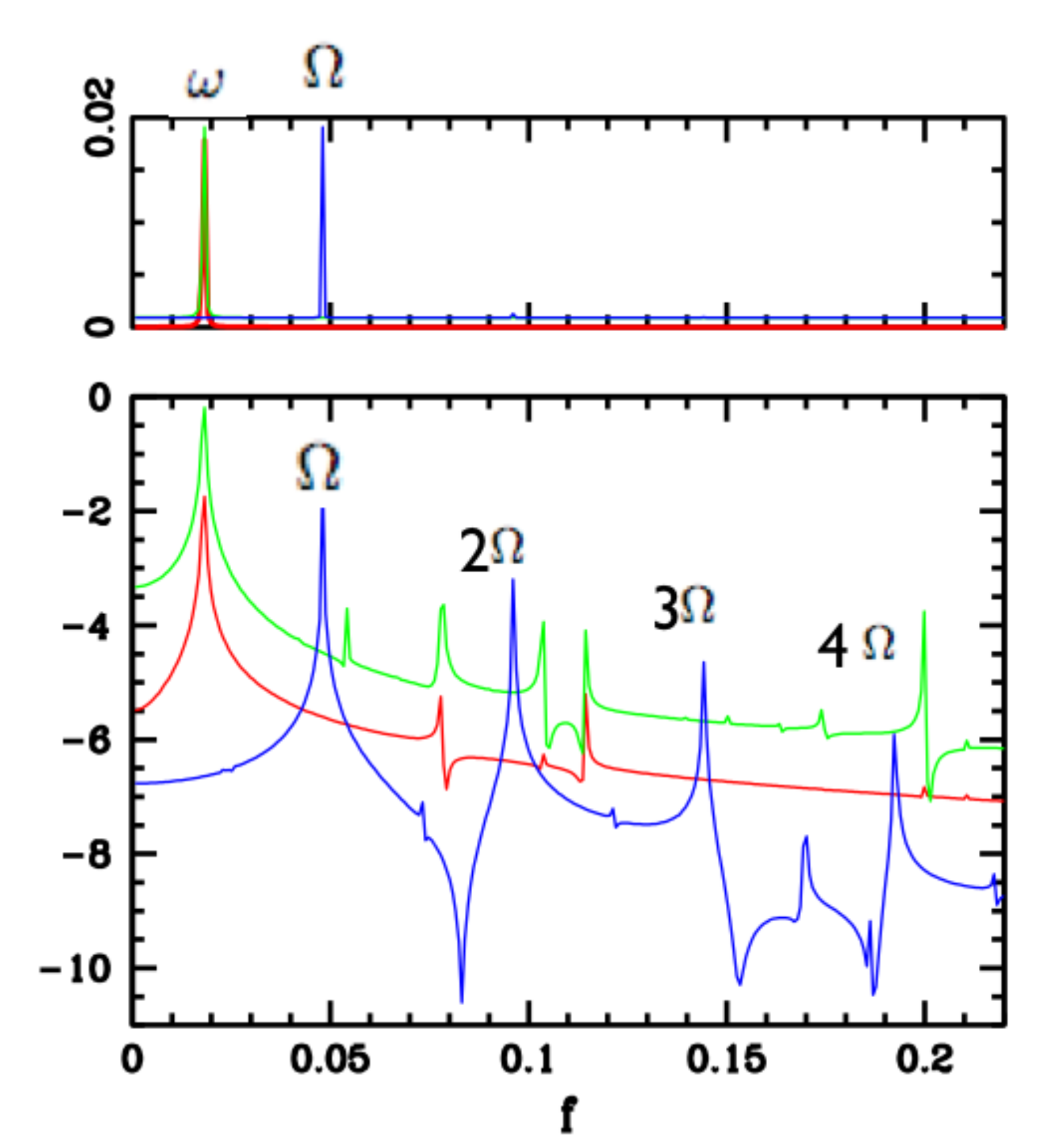}
\caption{(color online) Power spectrum or Fourier transform of the time
evolution of the spatially-averaged spin population $\rho_{m=+1}$ (red),
spatially-averaged spinor phase ${\gamma}$ (green), and spin-averaged
population $\sigma_{n=1}$ at site $n=1$ (blue).  The parameters and
initial state are as in Fig.~\ref{spin6} and the frequency is in units
of $J/\hbar$. The top and bottom panel show the power spectra on a linear
and logarithmic scale, respectively.  Also indicated are the dominant
frequencies $\omega$ and $\Omega$, which have been assigned as due to the
spin-dependent and spin-independent interactions, respectively. In the
bottom panel higher harmonics of $\Omega$ can be observed.}

\label{powerspin6}
\end{figure}

\section{Initial Discrete Soliton} \label{sec_initial}

For a ring lattice of six sites, we use the initial condition shown in
Fig.~\ref{F0}.  Commonly referred to as a discrete soliton, it is a
stationary solution of the DNLSE in the absence of the spin-changing
interaction \cite{Pando2004,Pando2005} (i.e. with $c_2=0$) and $c_0=J$.  Along with
phase $\phi_n^m=0$ or $\pi$ for even or odd site index $n$, respectively,
the initial state populations are the same for the three components.

This stationary solution, corresponding to a solution $\psi^m_n(t)=
\psi^m_n(t=0) \exp(-i\mu t/\hbar)$ where $\mu$ is the chemical potential,
is obtained by numerically solving the resulting nonlinear map, as
Eqs.~\ref{eq_one}-\ref{eq_three} reduce to a set of $L$ two-dimensional
cubic maps.  Following Ref.~\cite{Pando2009}, we find that for $L=6$ the
solutions are $6$-fold degenerate and for a range of $\mu$ the localized
soliton mode in Fig.~\ref{F0} is the only stable solution. In fact, we have used $\mu=2.5J$. Intriguingly,
for these values of the chemical potential the homogeneous solution with
the same density at all sites is unstable.

It should also be noted that known localized soliton-type solutions
with zero phase for all sites correspond to attractive spin-independent
interactions with $c_0<0$. To obtain localized solutions for repulsive
interactions, one needs to consider solutions with phases different
from zero.  In general, it can be shown that for a lattice with even
number of sites, the mapping $\psi^m_n \to   (-1)^n (\psi^m_n)^*$
relates a soliton solution for attractive interactions with those with
repulsive interactions.  For a single-component condensate discrete
solitons have been studied extensively for ring lattices of various
sizes \cite{Pando2004,Buonsante2007,Pando2009}.

Experiments with single-component Bose condensates in double-well
potentials \cite{Albiez2005} have observed Josephson oscillations and
quantum self-trapping in the limit $J\ll c_0$.  The opposite limit can
also be reached leading to tunneling of (nearly-)independent atoms. Here,
we chose a compromise with $c_0=J$.  As an aside we note that with our
initial state and $c_0=J$ we have implicitly specified the chemical
potential and thus atom number in each site.

\section{Quasiperiodic Dynamics of an anti-ferromagnetic spinor} \label{sec_weakspin}

Figure \ref{spin6} shows the time evolution of the spinor soliton on a
six-site ring, described in terms of the spinor coordinates $Z_n$ and $Z$,
spinor phases $\gamma_n$ and $\gamma$, as well as populations $\sigma_n$.
We use a small positive spin-changing interaction strength $c_2$ for an
anti-ferromagnetic spinor.  We observe that the time dependence of $Z$
and $\gamma$ are nearly sinusoidal. The site-dependent $Z_n$, $\gamma_n$
and $\sigma_n$, however, oscillate at a higher frequency. They do so in
a non-sinusoidal manner with sharper minima than maxima or vice versa.
The phases $\gamma_n$ only show small excursions around the average
$\gamma$.
Finally, we note that the spinor phases are bounded for oscillatory
motion.

Figure \ref{powerspin6} shows the power spectrum of three of the time
traces shown in Fig.~\ref{spin6}.  It highlights the existence of two
dominant frequencies, $\omega$ and $\Omega$, as well as weaker higher
harmonics in $\Omega$ indicating non-sinusoidal periodic behavior.
The spatially-averaged degrees of freedom predominantly oscillate with
a frequency $\omega$, which from simulations with other small $c_2$ is
found to be proportional to the absolute value of $c_2$.  In contrast,
spin-averaged local populations oscillate nearly-sinusoidal with frequency
$\Omega$, which from other simulations is found to weakly depend on the
spin-changing interaction but is inversely proportional to the number
of lattice sites.   Finally, the local dynamics for individual spin
projections is quasiperiodic with a slow frequency $\omega$ and a faster
beat frequency at multiples of $\Omega$. This illustrates the correlations
that exist among the spatial and spin components.  Detailed studies with
various ring sizes indicate that the spinor dynamics is characterized
by two frequencies, irrespective of the size of the ring.

\section{Resonance condition for the anti-ferromagnetic condensate} \label{sec_resonance}

\begin{figure}
\includegraphics[scale=0.42,trim=20 150 0 50, clip]{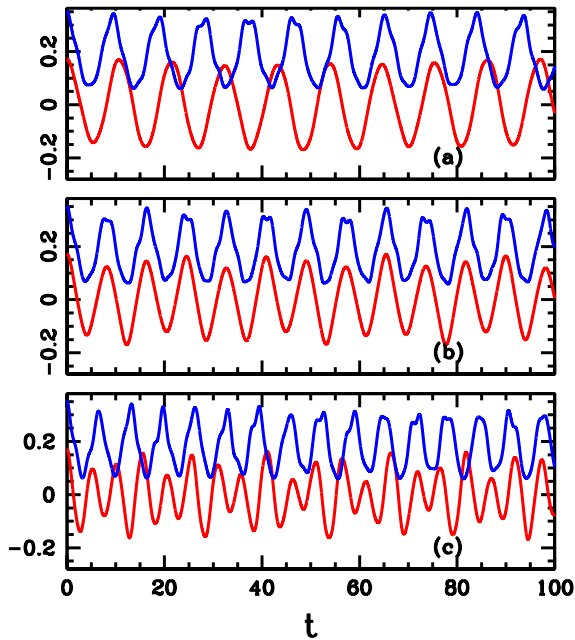}
\caption{(color online) 
Time series for an anti-ferromagnetic spinor showing the resonance
condition for increasing spin-changing strength $c_2$.  Time is in
units of $\hbar/J$ and $c_0=J$.  In all panels $Z$ (red curve) and
$\sigma_{n=1}$ (blue curve) are shown.  Panels a), b), and c) show
results for $c_2=0.5J$, $0.65J$, and $1.0J$, respectively.  The resonance
condition occurs for $c_2=0.65 J$, where the spatially-averaged $Z$ and
spin-averaged $\sigma_{n=1}$  oscillate at the same rate. For $c_2>0.65 J$
the traces are chaotic.}

\label{AFerrotime}
\end{figure}

\begin{figure}
\includegraphics[scale=0.4,trim=20 30 0 0,clip]{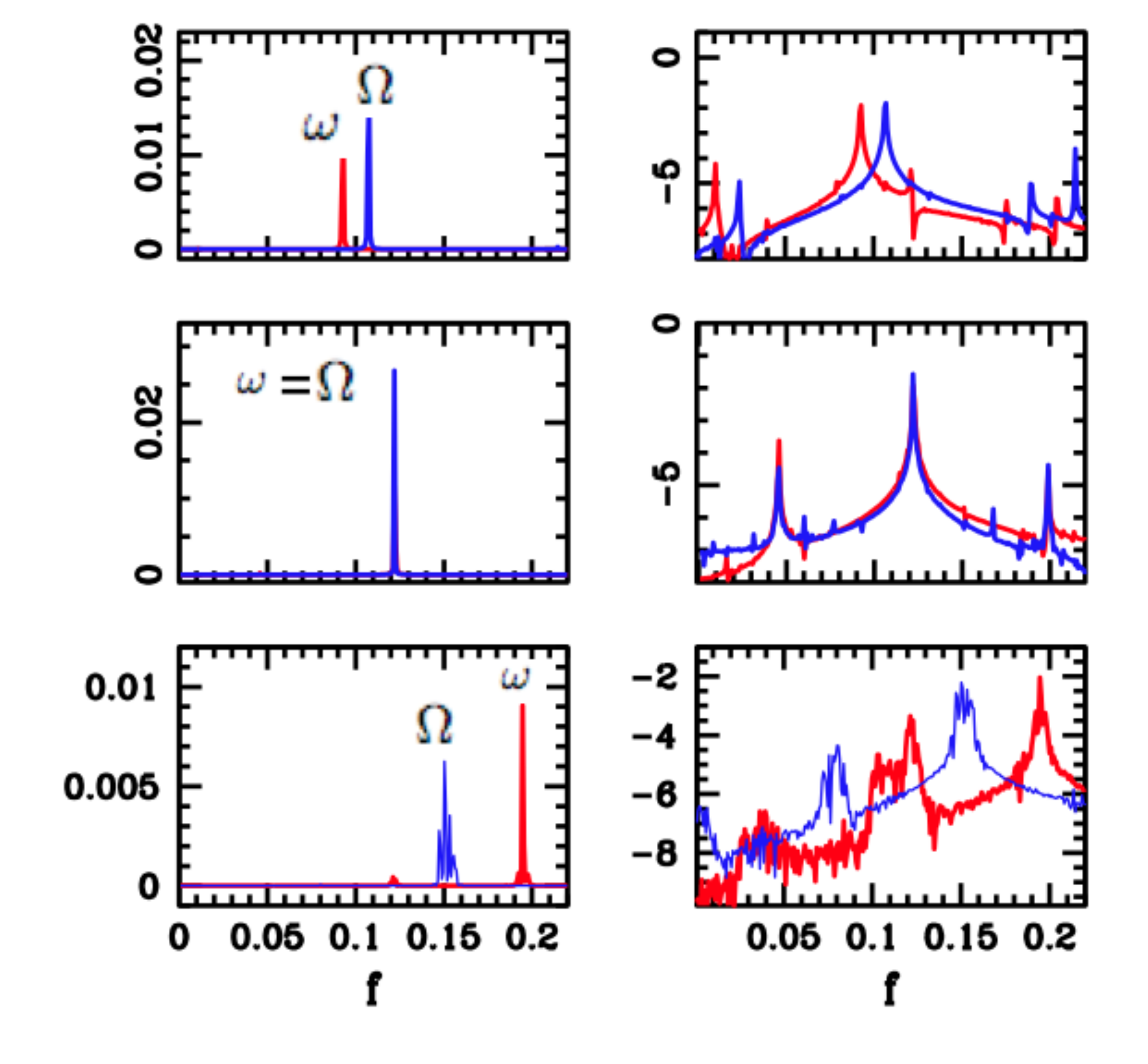}
\caption{(color online) 
Power spectra showing the resonance condition for increasing
spin-changing strength $c_2$ for the same parameters as in
Fig.~\ref{AFerrotime}.  The frequency is in units of $J/\hbar$.
In all panels the spatially-averaged $\rho_{m=+1}$
(red curve) and spin-averaged $\sigma_{n=1}$ (blue curve) are
shown. 
Panels on the left and right show spectra on a linear and logarithmic
scale, respectively, while from top to bottom $c_2=0.5J$, $0.65J$ and $1.0J$.}

\label{trans} 
\end{figure}

\begin{figure}
\includegraphics[scale=0.34,clip]{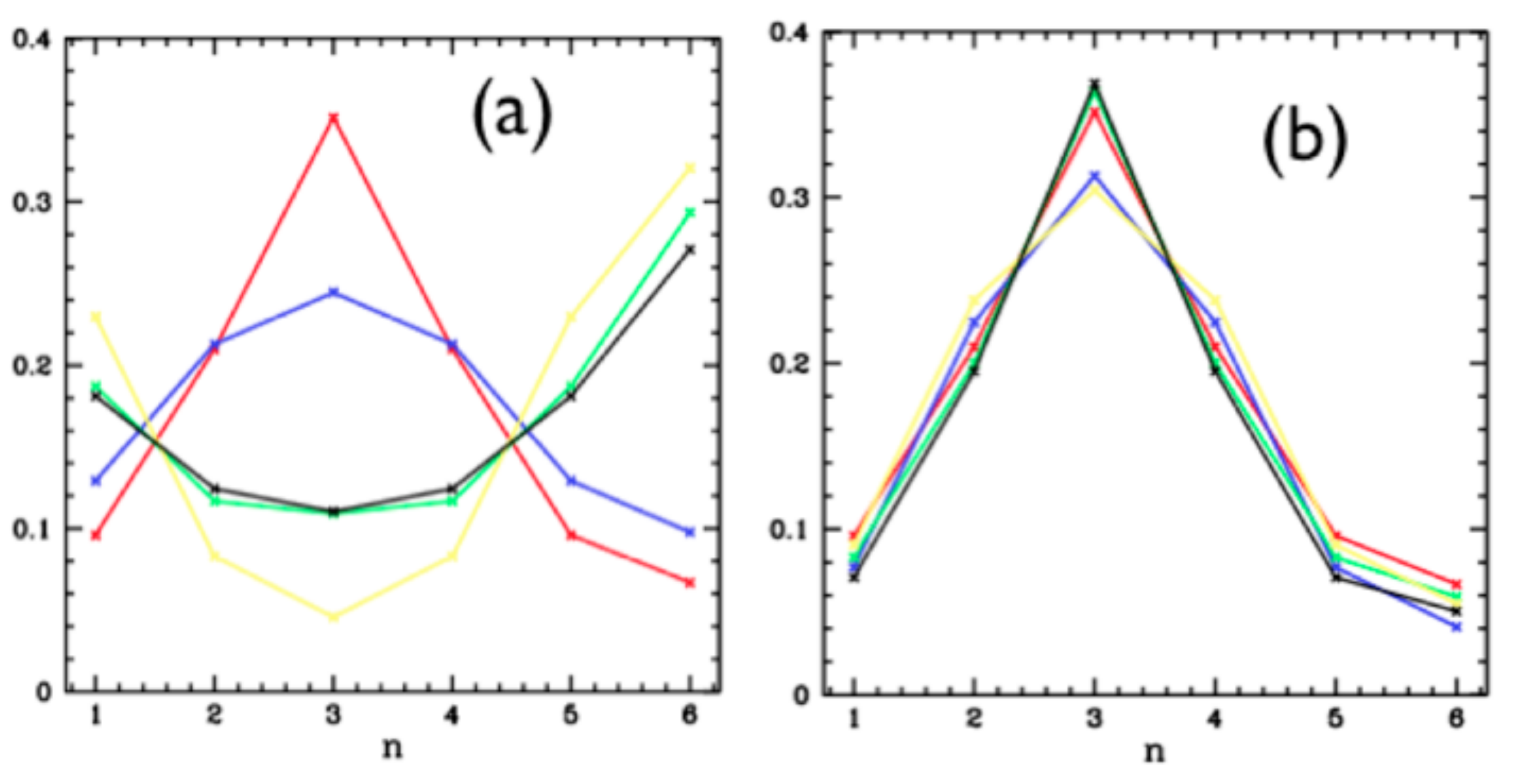}
\leavevmode \caption{(color online) Spin-averaged soliton profiles,
$\sigma_n$, for the anti-ferromagnetic spinor before and at resonance
for five times spaced by time interval $2\pi/\Omega$, starting with the initial state (red curve).  We use
$c_0=J$ and panels a) and b) correspond to $c_2=0.5J$ and $c_2=0.65J$,
respectively.  The soliton reforms close to its initial state at resonance
when $c_2=0.65J$.}

\label{Soltrans}
\end{figure}

Figures \ref{F0} and \ref{spin6} showed that for small $c_2/c_0$
two frequencies $\omega$ and $\Omega$ are very distinct.  For
increasing positive $c_2$ at fixed $c_0$, as shown by time traces in
Fig.~\ref{AFerrotime} and power spectra in Fig.~\ref{trans},  both
frequencies increase although at a different rate.  For a critical
spin-changing strength $c_2$ when $\omega=\Omega$ the anti-ferromagnetic
spinor reaches a ``resonance'' state.  All three
spin-components at all the sites of the ring then oscillate as a single
entity. For $c_0=J$ this resonance occurs
at $c_2=0.65J$.  For large values of $c_2$ the behavior becomes chaotic,
which is apparent as broad-band features in the corresponding power spectrum shown in Fig.~\ref{trans}.

Fourier analysis of collective as well as local coordinates shows that
the resonance state corresponds to matching of not only the dominant
frequencies  $\omega$ and $\Omega$, but also some of the other less
prominent frequencies (not visible in the linear plot) as illustrated in
the log-linear plot on the middle row of Fig.~\ref{trans}.  In fact, at
resonance, the dominant or the primary peak is accompanied by secondary
satellite peaks, equally spaced on the either side of the primary peak.

Figure \ref{Soltrans} further illustrates that at resonance all sites
oscillate in phase with same frequency.  The spatial profile of the
soliton reemerges periodically without any significant change from the
initial profile.  Thus, the resonance state in an ordered state where
wave functions of all the components of the spinor condensate at all
$L$-sites of the ring lattice oscillate in unison and the dynamics is well
characterized one frequency.  The fact that the soliton profile reemerges
periodically provides a unique demonstration of quantum coherence and
a novel-type of internal spatio-temporal Josephson effect.

\begin{figure}
\includegraphics[scale=0.43,trim=20 150 0 70, clip]{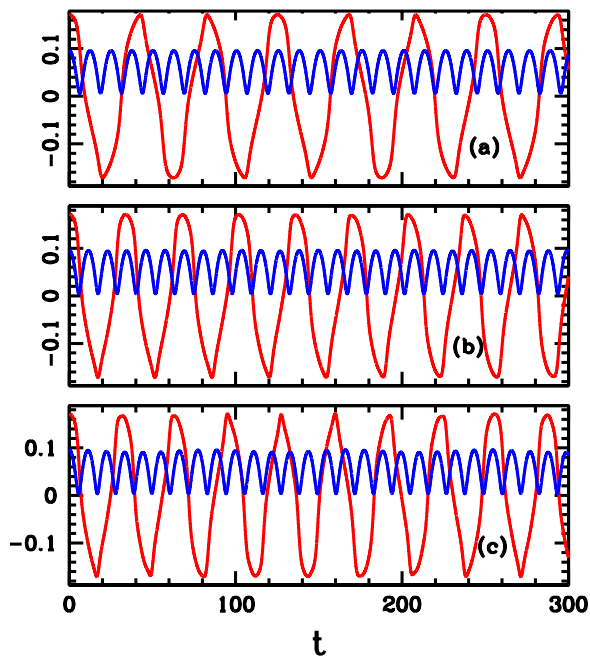}
\caption{(color online) Time evolution of spatially-averaged population
$Z$ (red) and spin-averaged population $\sigma_{n=1}$ (blue) for a
ferro-magnetic spinor condensate with negative $c_2$.  Panel a), b), and
c) show traces for $c_2=-0.075J$, $-0.09J$, and $-0.095J$, respectively,
corresponding to cases before, at, and beyond resonance.  We use $c_0=J$
and time is in units of $\hbar/J$.}

\label{Ferrotime}
\end{figure}

\begin{figure}
\includegraphics[scale=0.43,trim=0 170 0 80,clip]{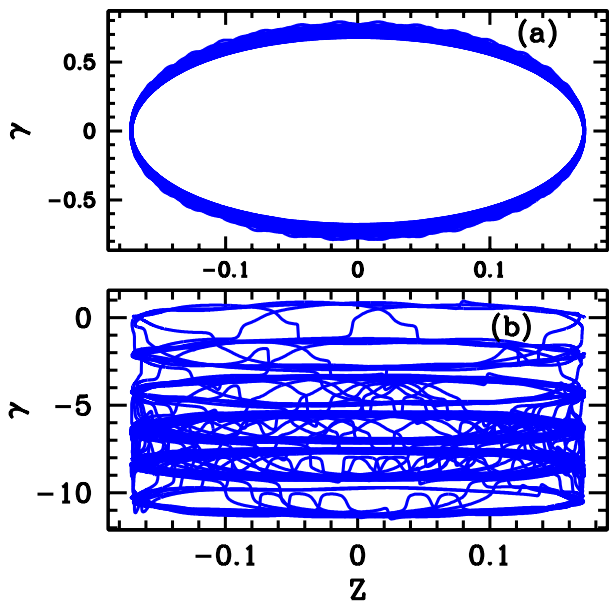}
\caption{Phase portraits or parametric plots of the time-evolution of the
spinor population $Z$ and phase $\gamma$ for a ferromagnetic spinor at
$c_2=-0.075J$, panel a), and  $c_2=-0.095J$, panel b).  The system is at
resonance in panel b). 
The dynamics in panel b) is accompanied by unbounded motion of the
spinor phase. We use $c_0=J$.}

\label{ppFerro}
\end{figure}

\section{ Ferromagnetic Condensate}\label{sec_ferro}

We now briefly describe the dynamics of a ferromagnetic spinor
condensate with negative $c_2$. For a small spin-changing interaction,
the dynamics is similar to that of an anti-ferromagnetic spinor. Namely,
the spatially-averaged behavior is pendulum-like with dominant frequency
$\omega$, while the local population oscillates with frequency $\Omega$.
An example is shown in Fig.~\ref{Ferrotime}. The local density for each
spin component oscillates with both frequencies.  A resonance state can
again be achieved by increasing $|c_2|$ but, for $c_0=J$, now occurs when
$\omega/\Omega = 3$.

In contrast to the anti-ferromagnetic case, however, the resonant
transition is accompanied by unbounded dynamics or phase winding where the
phase of the condensate becomes unbounded as shown in Fig.~\ref{ppFerro}.
This behavior manifests itself as a zero-frequency mode in a power
spectrum. In other words, chaotic dynamics with its broad-band
spectrum is accompanied by a transition from an oscillatory to rotational
mode for the collective degrees of freedom.

\section{ Discussion}\label{sec_discussion}

In summary, numerical explorations of mean-field equations of spinor
condensate on a ring lattice with a small number of sites reveals
strikingly correlated dynamics of the many-body system. Even though the
condensate is separated into $L$-sites, soliton dynamics at all sites can be
characterized by just two frequencies.  With quasiperiodic dynamics at
local sites for individual spin components, spatially averaged behavior
for each spin component as well as spin-averaged dynamics at each sites
is found to be periodic. The fact that the time series describing
local dynamics is characterized by two frequencies and these two
frequencies untangle in the collective degrees of freedom is rooted in
correlations among different spin degrees of freedom at various sites of
the lattice. However, proper understanding of these correlations remains
an open challenge.  Our study with ring lattices of various sizes show
that two-frequency characterizations of the spin-$1$ condensate on a ring
lattice is valid irrespective of the number of sites on the lattice.

Our study provides a new illustration of a quasiperiodic route to chaotic
dynamics in a many-body system where the critical point is known to be
characterized by a resonant state.  Simple models of dynamical systems such
as the one-dimensional circle map\cite{Ott1993} are paradigms of quasiperiodic
route to chaos, where the critical point corresponds to parameters where
the two frequencies are mode-locked. The quasiperiodic route to chaos is
a well-established scenario in dynamical systems exhibiting a transition
from regular to chaotic dynamics \cite{Ott1993}. However, the fact that
the critical point describes periodic dynamics is a novelty in many-body
systems rooted in the coherence associated with spinor condensate.
In this case, the critical point is a highly-ordered many-body state
exhibiting a new type of spatial-temporal coherence.  At the critical
point where the two dominant frequencies are in resonance, the ring
lattice with $L$-sites oscillates in unison with a single characteristic
frequency. This internal Josephson effect where an unscathed soliton profile
reemerges periodically provides a novel illustration of both spatial
and temporal coherence.

This research is supported by Office of Naval Research,
the CONACYT-M\'{e}xico and
the US Army Research Office.

\bibliography{refs}

\begin{thebibliography}{32}
\bibitem[{sol()}]{soliton}
\bibinfo{note}{A soliton is a solitary wave with a special collision property,
  viz., it retains its identity even after collision with another solitary
  wave. Following a common practice in physics lierature, we will refer to
  nonlinear localized solitary waves as solitions here.}
\end{thebibliography}
\end{document}